\documentclass[fleqn,10pt]{wlscirep}
\title{Graphlets in Multiplex Networks}

\usepackage{mathptmx}
\usepackage{array}
\usepackage{moreverb}
\usepackage{mathtools}
\DeclarePairedDelimiter\floor{\lfloor}{\rfloor}

\author[1,3,+]{Tamara Dimitrova}
\author[1,3,+]{Kristijan Petrovski}
\author[1,2,3,*]{Ljupco Kocarev}
\affil[1]{Macedonian Academy of Sciences and Arts, Skopje, Republic of Macedonia}
\affil[2]{Faculty of Computer Science and Engineering, UKIM, Skopje, Republic of Macedonia} 
\affil[3]{IKT-Labs, Skopje, Macedonia}

\affil[*]{lkocarev@manu.edu.mk}

\affil[+]{These authors contributed equally to this work}

\keywords{Graphlets, Multiplex Networks, Social Networks, Economy Trade Networks}  

\begin{abstract}
 
We develop  graphlet analysis for multiplex networks and discuss how this analysis can be extended to multilayer and multilevel networks as well as to graphs with node and/or link categorical attributes.   The analysis has been adapted for two typical examples of multiplexes: economic trade data represented as a 957-plex network and 75 social networks each represented as a 12-plex network. We show that wedges (open triads) occur more often in economic trade networks than in social networks, indicating the tendency of a country to produce/trade of a product in local structure of triads which are not closed.  Moreover, our analysis provides evidence that the countries with small diversity tend to form correlated triangles. Wedges also appear in the social networks, however the dominant graphlets in social networks are triangles (closed triads). If a  multiplex structure indicates  a strong tie, the graphlet analysis provides another evidence for the concepts of strong/weak ties and structural holes. In contrast to Granovetter’s seminal work on the strength of weak ties, in which it has been documented that the wedges with only strong ties are absent, here we show that for the analyzed 75 social networks, the wedges with only strong ties are not only present but also significantly correlated.

\end{abstract}
\begin{document}

\flushbottom
\maketitle

\thispagestyle{empty}

\section*{Introduction}

Networks are ubiquitous. This has triggered a vast amount of research in the last two decades. When focusing on the local connectivity of subgraphs within a network, two approaches have been identified, motifs and graphlets. Motifs are defined as sub-graphs that repeat at frequency higher than in the random graphs \cite{milo2002network,sporns2004motifs}. However, they depend on the choice of the network's null model. In contrast, graphlets~\cite{prvzulj2004modeling} are induced sub-graphs of a network that appear at any frequency and hence are independent of the null model. Graphlets have found numerous applications as building blocks of network analysis in various disciplines ranging from social science \cite{holland1976local,faust2010puzzle} to biology \cite{prvzulj2007biological,yaverouglu2014revealing}. In social science, graphlet analysis (known as sub-graph census) is widely adopted in sociometric studies \cite{holland1976local}. Recently, graphlet analysis has also been adapted for directed networks \cite{sarajlic2016graphlet,trpevski2016graphlet}. 

Real networks come in various structures including multilayer networks, multiplex networks, multilevel networks or multiscale networks. Understanding local connectivity of subgraphs by using graphlets in these networks could, therefore, lead to improved predictive modes of networks and/or to enhanced description of their properties. In sociology, the importance of multiplex networks has been emphasized by many scholars. White, Boorman and Breiger  \cite{Boorman1976} and Boorman and White \cite{White1976} treated multiple networks as a foundation of social structure and argued that the patterning and interweaving of different types of ties are needed to describe and characterize social structures. Multiplexity is critical to diverse phenomena such as the mobilization of social movements \cite{gould1991multiple}, the consolidation of political power \cite{Padgett1993}, the emergence of trust in economic relationships \cite{granovetter1985economic}, the creation of social bonds within civic networks \cite{baldassarri2007integrative}, and the organization of party coalitions \cite{Grossman2009}. Multiplex networks have been studied to understand scientific collaboration \cite{Maggioni2013}, structural logic of intra-organizational networks \cite{Rank2010}, formation of ties featuring both an economic and a social component in inter-organizational networks \cite{Ferriani2012}, and formation of relationships among producers in the multiplex triads \cite{Shipilov2012}.

Multilayer/multiplex networks \cite{de2013mathematical,kivela2014multilayer} have recently been a subject of particularly intense research by the network science and physics communities. Novel structural descriptors \cite{sola2013eigenvector,halu2013multiplex,battiston2014structural,de2015ranking,cozzo2015structure}  and tools from statistical physics \cite{bianconi2013statistical,menichetti2014weighted}  have been developed for studying multilayer networks. By analyzing multilayer networks, instead of relying on their monolayer counterparts, scientists have documented evidences of novel features and novel insights about real systems (see, for example, \cite{cardillo2013emergence,braun2015dynamic,filiposka2017bridging}). Multilayer networks are fundamental for understanding of dynamical processes on networked systems, including, for example, spreading processes, such as flows (and congestion) in transportation networks \cite{morris2012transport,sole2016congestion} , and information and disease spreading in social networks \cite{wang2015coupled,funk2015nine,granell2013dynamical,sanz2014dynamics,lima2015disease}.

This paper aims at developing graphlet analysis for multiplex networks. The study has been motivated by two facts: (1) graphlets are powerful tool for analyzing single plex/layer networks, and (2) networks with multiplex/multilayer structures are common in nature and societies. 
The central problem in addressing graphlets in a multiplex network comes from the explosive growth in the number of various edge types with the linear increase in the number of plexes in the network. We now present an illustrative example showing graphlet analysis's challenges and how they are addressed for multiplex networks.

%
%
%

\begin{figure}
\centering
\includegraphics[width=\textwidth]{multiplex_example_full_compressed.pdf}
\caption[LoF entry]{ Upper-left panel: single plex graph. Nodes of a graphlet are classified into different orbits. 

Upper-right panel: multiplex graph with 3 plexes $a, b,$ and $c$. On the same panel the multiplex graph is represented as edge-labeled graph: each edge is labeled with a single label, an element of the set $\{a,b,c,ab,ac,abc\}$. 
   
Left table: triangles and wedge stars. Right table: wedge paths. The column `nodes' shows the nodes of the corresponding orbit (triangle, wedge star, or wedge path). The node in the small bracket (for the first triangle this is the node 1 shown as (1)) indicates the first node of the orbit. The column `edges' shows the full description of sub-orbits. While for the single plex graph (upper-left panel) all triangles are same, for the multiplex graph (upper-right pane) these triangles are different and are labeled with a single label from the set $E_t$. For example the edges of first triangle are labeled with $ab.ab.b$.  

The column `edges\_1' shows the first reduction of the number of sub-orbits. Thus, for example, all wedge stars $ab.ac$, $ab.ab$, $ab.bc$, and $ac.bc$ are represented as $2.2$. The column `edges\_2' represents another reduction in which $ab.ac$, $ab.bc$, and $ac.bc$ are represented as $2_x.2_y$, while $ab.ab$ is represented as $2_x.2_x$.

The number of wedge paths are doubled with paths in which the other non-central node is a starting node of the path.  } 
\label{fig:example_img}
\end{figure}

Consider a simple undirected (single-plex) graph $G=(V,E)$, see Fig.~\ref{fig:example_img}, for which 
\begin{eqnarray}
V &=& \{1,2,3,4,5,6,7,8\} \\ 
E &=& \{12,13,23,28,34,36,38,45,67,68,78\}.
\end{eqnarray}
A graphlet $G' = (V', E')$ is an induced subgraph of $G$ (see Materials and Methods). Thus, for example, $G' = (\{3,4,5\} , \{34,45\})$ is a graplet called wedge.    
By taking into account the ``symmetries'' between nodes in a graphlet, the nodes of the graphlet can be classified into different orbits (see Materials and Methods). The node 3 of the graphlet $G' = (\{3,4,5\} , \{34,45\})$  belongs to the orbit 1 called ``wedge star'', while the node 4 of the graphlet $G'$ is classified as ``wedge path'' and is denoted as orbit 2. The orbit 0 of a graphlet is node degree.      
Consider now a 3-plex network shown on Figure \ref{fig:example_img}. We write  $a,b,c$ for network plexes. Let $E_t = \{a,b,c,ab,ac,bc,abc\}$.       
A multiplex graph is a special type of edge-labeled graph in which each edge is labeled only with a \textit{single} label -- an element of the set $E_t$. Thus, for our example, the multiplex graph can be written as labeled $G=(V,E_{label})$, see Fig. \ref{fig:example_img}, for which 
\begin{eqnarray}
V &=& \{1,2,3,4,5,6,7,8\} \\ 
E_{label} &=& \{12_{ab},13_{ab},23_b,28_b,34_{ac},36_{ab},38_{bc},45_b,67_{abc},68_{c},78_b\},
\end{eqnarray}
where $e_\alpha$ means the edge $e \in E$ is labeled with $\alpha \in E_t$ .   
Two problems will be addressed when dealing with graphlet analysis of multiplex networks. The first problem is related to the fact that each orbit of a multiplex network consists of sub-orbits. Thus, the first orbit, the orbit 0, consists of 7 sub-orbits: $0_a$, $0_b$, $0_c$, $0_{ab}$, $0_{ac}$, $0_{bc}$, and $0_{abc}$.
To successfully define the sub-orbits for other orbits (wedges, triangles, and so on) we first introduce two notations: (1) permutation $[a,b,c]$, stressing the order of edge types as part of the graphlet and (2) set $\lbrace a,b,c \rbrace$  stressing the symmetry the edges have as part of the graphlet.
Thus, for example, the sub-orbit of the orbit wedge star for the node 4 in the graphlet with vertex set $\{3,4,5\}$ is $2_{ac.b}$ which is for simplicity on the Figure 1 written as $ac.b$. For our example, all sub-orbits of the orbit 2 (wedge star) are elements of the set $2_{\{\alpha.\beta\}}$ where $\alpha, \beta \in E_t$; the cardinality of this set (the set of wedge star sub-orbits) is 28.     
The sub-orbit of the orbit wedge path for the node 2 in the graphlet with vertex set $\{2,8,6\}$ is $1_{b.c}$ or simply $b.c$.  For our example, all sub-orbits of the orbit 1 (wedge path) are elements of the set $2_{ [\alpha.\beta] }$ where $\alpha, \beta \in E_t$; the total number of wedge star sub-orbits is 49. Section Materials and Methods describes how sub-orbits are defined for a given orbit and the size of the orbit (the cardinality of the set of sub-orbits associated with the orbit). 

The second problem is the size of the orbit. Even for the simplest graphlet, a node degree, in a multiplex network the size of orbit 0 (node degree) equals $2^d -1$, where $d$ is the number of plexes. 
This limits the application of graphlets for real data even for networks with small number of plexes.  In order to address this problem, we propose two  different methods for reducing the number of graphlets. 
Both methods are illustrated on Figure \ref{fig:example_img}.  
The simplest possible reduction of the  $E_t = \{ a,b,c ,ab,ac, cb,abc \}$ is to the set $E_t^{'}=\{1,2,3 \}$. In another words, an edge of a multiplex, which has been defined as an element of the set $E_t$ representing different relation types, is now (after reduction) defined as an element of the set $E_t^{'}$ representing the `strength' (plex count) of the original edge. Thus wedge stars for nodes 3,1,4 and 3,1,6 which represent two different sub-orbits $ab.ac$ and $ab.ab$, respectively, are now represented as a single sub-orbit $2.2$.  
More generally, in this example, all sub-orbits 
$\alpha.\beta$ such that $\alpha,\beta \in E_t$ and $|\alpha| = |\beta|=2$ are represented as a single orbit  $2.2$ (see Figure 1).   
The second reduction takes into account that the  sub-orbit $2.2$ is result of reduction of two subsets: 
\begin{eqnarray}
A &=& \left\{ \alpha.\beta: \alpha \neq \beta, \alpha,\beta \in E_t,  |\alpha| = |\beta|=2 \right\} \\ 
B &=& \left\{ \alpha.\alpha: \alpha \in E_t, |\alpha|=2 \right\}.  
\end{eqnarray}
Both subsets $A$ and $B$, in this reduction, are mapped to two different sub-orbits $2_x.2_y$ and $2_x.2_x$, respectively (see Figure \ref{fig:example_img}).      
This reduction results in a multiplex which is  called plex-count multiplex with distinct links inside orbits. Section Materials and Methods provides the full description of both reductions.

The graphlet analysis for multiplex networks can be extended to graphs with node and/or link (categorical) attributes as well as to multilayer and/or multilevel networks as discussed in the supporting information (SI).  Further, economic and social multiplex networks are analyzed using graphlets providing novel insight of networks’ properties and/or their local structures. In particular, for economic networks we show that (1) countries produce/trade products in local structure of triads which are not closed and (2) counties with small diversity tend to form correlated triangles. For social networks for which a strong tie is related to the multiplex structure, we provide an example of social networks for which the wedges with only strong ties are both present and significantly correlated, in contrast to the Granovetter’s seminal work on the strength of weak ties, in which it has been shown that the wedges with only strong ties are absent.


\section*{Materials and Methods}

\subsection*{Multilayer/multiplex networks and graphlets}

This paper introduces a method for graphlet analysis of multiplex networks. This analysis can also be adopted for graphs with node and/or edge (categorical) attributes. In the SI we explain how the method can be generalized to graphs with node and/or edge attribute and complex (multiplex, multilayer, and multilevel) graphs. Next, definitions of multiplex, multilayer and multilevel networks as well as graphlets for simple graphs are provided.

\noindent
\textit{Multiplex networks} --   A multiplex network is defined as a  $d+1$-tuple $G = (V, E^1, \ldots, E^d)$  where  $V$ is the set of nodes  and for each $\alpha \in \{1, 2, \ldots, d \}$, $E^\alpha$ is the set of edges describing the presence or absence of edges of type $\alpha$ between pairs of nodes. Since a multiplex network is uniquely defined with the node set $V$ and the edge sets $E^1, \ldots, E^d$, we write $G(V, E^1, \ldots, E^d)$ to denote the multiplex.  The graph $(V,E^\alpha)$ is also called plex. A $k$-plex network is a subgraph of the multiplex and is defined as $G=(V, E^{\alpha_1}, \ldots, E^{\alpha_k})$.

\noindent
\textit{Multilayer networks} -- A multilayer network is defined as a graph $G = (V,E)$ for which $V \subseteq V^1 \times V^2 \ldots \times V^d$ and $E$ is the set of edges. Typically, $V^\alpha = V$, $\alpha = 1, \ldots, d$; the elements of $V$  are called nodes. A layer is a sub-graph  $(V^\alpha, E^\alpha)$ for which $E^\alpha = \{ ij \in E: i, j \in V^\alpha \}$. Let $A = [a_{ij}^{\alpha \beta}]$, $i,j =1, \ldots n$ and $\alpha, \beta = 1 \ldots m$, be the adjacency matrix of the graph $G = (V, E)$. In general it is assumed that $a_{ij}^{\alpha \beta} \neq 0$ for $\alpha \neq \beta$ and $i \neq j$. In the special case when a network is represented with an adjacency matrix such that $a_{ij}^{\alpha \beta} = 0$ when both $\alpha \neq  \beta$ and $i \neq j$, the network is also called multiplex network. This definition implies that a multiplex consists of layers (not plexes). In what follows, we will not consider multilayer approach of multiplex networks.   

\noindent
\textit{Multilevel networks} -- A multilevel (interconnected) network is a graph $G = (V,E)$ for which $V = \cup^d_{\alpha=1}V^\alpha$ and $E$ is the set of edges. In general, $V^\alpha \neq V^\beta$ for $\alpha \neq \beta$; each $V^\alpha$ represents a distinct type of nodes. Multilevel networks can be described with subgraphs $(V^\alpha, E^\alpha)$ called levels, where $E^\alpha = \{ ij \in E: i, j \in V^\alpha \}$ and bipartite subgraphs $(V^\alpha, V^\beta, E^{\alpha \beta})$ such that  $E^{\alpha \beta} = \{ ij \in E: i \in V^\alpha, j \in V^\beta \}$.    


%


\noindent
\textit{Graphs with attributes} --
Let $G = (V, E)$ be a simple graph,  $V_A$ be the set of node attributes, and $E_A$ be the set of edge attributes. Nodes are labeled $i = 1, 2, \ldots, |V|$, node attributes are labeled $q = 1, 2, \ldots,  |V_A|$, while edge attributes are labeled $\alpha = 1, 2, \ldots, |E_A|$. 
We define $ |V| \times |V_A|$ matrix $D = [d_{iq}]$ as $d_{iq} = 1$ if and only if the node $i$ has the attribute $q$, otherwise 0. 
We define a sub-graph $G^\alpha = (V, E^\alpha)$ as $E^\alpha = \{ e | e\in E \mbox{ and } e \mbox{ has attribute } \alpha \}$. 

\noindent
\textit{Graphlets} --
Let $G = (V,E)$ be a graph, where $V$ is a set of nodes and $E$ is a set of edges. A subgraph $G'$ of $G$ is a graph whose set of nodes and set of edges are subsets of $G$. An induced subgraph of $G$, $G' = (V', E')$, is a subgraph that consists of a subset of nodes in $G$ and all of the edges that connect them in $G$, i.e. $V' \subset V$, $E' = \{(u, v) : u, v \in V \}, (u, v) \in E'$. The size of a graphlet is the cardinality of its node set. Unless we explicitly say ``induced'' in this paper, a subgraph is not necessarily induced. The top part of Table \ref{tab:explain} shows all 2-, 3-, and 4-node undirected graphlets $G_i$, 0 $\leq i \leq 8$. By taking into account the ``symmetries'' between nodes in the graphlet $G_i$, the nodes of $G_i$ are classified into different automorphism orbits (or just orbits, for brevity), where the nodes with the same orbit identification are topologically identical. For all $G_i$, $0 \leq i  \leq 8$, there are 15 orbits, which are also shown as filled nodes in the top panel of Table~\ref{tab:explain}.

\subsection*{Graphlets in multiplex graphs}



Table \ref{tab:explain} depicts all 2-4 node graphlets with their orbits (15 in total) for a simple graph $G=(V, E)$. For multiplex graphs, these orbits are further subdivided based on the relation (edge) types, so that each orbit consists of a number of \textit{sub-orbits} that are defined by the specific edge types inside the graphlet. 
Let $ {\cal R} = \{1, 2 , \ldots, d \}$ be the set of relation/edge types.  Let ${\cal P}({ \cal R })$ be the power set of the set ${\cal R}$, that is, the set of all subsets of ${\cal R}$, we define $E_t = {\cal P }( {\cal R}) \setminus \emptyset $. The set $E_t$ is the set of combinations of different edge types with cardinality $| E_t | = 2^d -1 $. 
Multiplex graphs are treated as having $2^d-1$ different types of edges so that each edge of a multiplex is uniquely defined with an element of the set $E_t$. Thus, for example, for $ {\cal R} = \{1, 2 ,3 \}$, an edge of the multiplex is represented by an element of the set $\{  1, 2, 3, 12, 13,23, 123  \}$. To simplify the notation, the set $\{a,b,c\}$ - the element of the set $E_t$ - will be labeled as $abc$. Thus, the first orbit – the orbit 0 – see Table \ref{tab:explain}, consists of 7 sub-orbits: $0_1$, $0_2$, $0_3$, $0_{12}$, $0_{13}$, $0_{23}$, and $0_{123}$.       
To successfully define the \textit{orbit class} for each orbit we first define/introduce two notations: (1) permutation $[a,b,c]$, highlighting the order of the types of edges as part of a graphlet and (2) set $\lbrace a,b,c \rbrace$ (can have repeating elements, only their order does not matter), highlighting the symmetry the edges have as part of a graphlet. Thus for example, in a multiplex the orbit 1 consists of the orbit class $1_{[a,b]}, a,b \in E_t$ while the orbit 2 consists of the orbit class $2_{\{a,b\}}, a, b \in E_t$. For  $ {\cal R} = \{1, 2\}$ the sub-orbit instances of orbit 1 are: $ 1_{1.1}$, $1_{1.2}$, $1_{1.12}$, $1_{2.1}$, $1_{2.2}$, $1_{2.12}$, $1_{12.1}$, $1_{12.2}$, and $1_{12.12}$, while the sub-orbit instances of orbit 2 are: $2_{1.1}$, $2_{1.2}$, $2_{1.12}$, $2_{2.2}$, $2_{2.12}$, and $2_{12.12}$. Lexical sorting (smaller plex count links first, lexical sort on same plex count links) is used to order the edges inside every wedge star sub-orbit. In this way every sub-orbit has an unique representation with which it is further identified. 

\begin{figure}
    \centering
    \includegraphics[width=\textwidth]{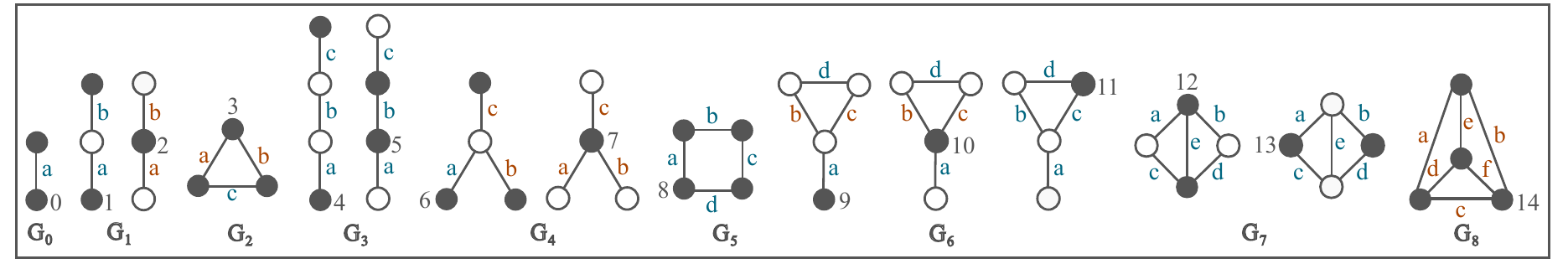}
    \caption{Sub-orbit breakdown for graphlets up to order 4. The top panel shows all multiplex $G_k$ graphlets, $0 \leq k \leq 8$ and their 15 orbits. The orbit nodes are colored gray. Every edge is identified by a letter. Each blue colored edge belongs to a permutation group ([]) inside the orbit class, while each orange edge belongs to a symmetric set group (\{\}). Edges that belong to symmetric sets are always lexically sorted (smaller plex count links first, lexical sort on same size links) before assigning them a specific sub-orbit instance.}
    \label{fig:suborbits}
\end{figure}

A straightforward way to enumerate all possible graphlet sub-orbits is described in the supplementary material. Table \ref{tab:explain} visually shows the first 15 orbits and gives information about their multiplex classes. Each orbit is uniquely represented with an orbit class, labeled as shown in Table \ref{tab:explain} and the cardinality of this class is called the size of the orbit. To simplify the notation we write $e^{O_i}_1.e^{O_i}_2. \cdots e^{O_i}_k$ for a sub-orbit of the given orbit $O_i$. When it is clear from the text what orbit is considered, we will drop the explicit $O_i$ in order to keep the notation uncluttered. 
Each orbit $O_i$ can then viewed as a set of sub-orbits, $O_i=\{ e_1.e_2.e_3..e_k  \hspace{0.1cm}| \hspace{0.1cm} e_j \in E_t\}$. Inside the orbit every list of edges $e_1..e_k$ is connected in the same way, however they are differentiated by the specific values of the multiplex links $e_j$ coming from the set $E_t$, explained visually in \ref{fig:suborbits}.

\begin{center}
$$
\begin{array}[H]{lllll}
\mbox{Orbit} & \mbox{Orbit Alias} & \mbox{Orbit class} &  & \mbox{Size of orbit} \\ \hline \\
0& \mbox{Degree} &{0_{[a]}} & {a \in {E_t}} &  {|E_t|} \\
1& \mbox{Wedge Path} &    {1_{[a, b]}} &  {a, b \in {E_t}} &  {|E_t|^2}\\
2& \mbox{Wedge Star}    &    {2_{\{a, b\}}} & {a, b \in {E_t}}& \left({{{|E_t|}\choose {2}}}\right)  \\
3& \mbox{Triangle} &{3_{[\{a, b\}, c]}} &  {a, b, c \in {E_t}} &  {\left({{{|E_t|}\choose {2}}}\right) \cdot |E_t|}\\
4  & \mbox{Four Path} &{4_{[a, b, c]}}  & {a, b, c \in {E_t}} & {|E_t|^3}\\
5&&{5_{{}[a, [b,c]]}}  & {a, b, c \in {E_t}}& {|E_t|^3} \\
6&\mbox{Three Star}&{6_{[a, \{b, c\}]}} &  {a, b, c \in {E_t}} & {|E_t| \cdot \left({{{|E_t|}\choose {2}}}\right)} \\
7&&{7_{\{a, b, c\}}}  & {a, b, c \in {E_t}} & {\left({{{|E_t|}\choose {3}}}\right)}\\
8&\mbox{Four Cycle}&{8_{\{[a, d],[b,c]\}}}  & {a, b, c \in {E_t}} & \left({{{|E_t|^2}\choose {2}}}\right) \\
9&\mbox{Tailed Triangle}&{9_{[a, \{b, c\}, d]}}  & {a, b, c, d \in {E_t}} & {|E_t| \cdot {{|E_t|}\choose {2}} \cdot |E_t|}\\
10&&{10_{[a, \{b,c\}, d]}} & {a, b, c, d \in {E_t}} & {|E_t| \cdot {{|E_t|}\choose {2}} \cdot |E_t|}\\
11&&{11_{[a,b,c,d]}}  & {a, b, c, d \in {E_t}} &  { |E_t|^4}\\
12& \mbox{Chordal Cycle} &{12_{[\{[a, c], [b,d]\}, e]}}  & {a, b, c, d \in {E_t}}  & \left({{{|E_t|^2}\choose {2}}}\right) \cdot |E_t| \\
13&&{13_{[\{[a,b], [c,d]\}, e]}}  & {a, b, c, d \in {E_t}}  &   \left({{{|E_t|^2}\choose {2}}}\right) \cdot |E_t|\\

14&\mbox{Four Clique}&{14_{\{a, b, c, d ,e ,f\}}}
& {a, b, c, d,e, f \in {E_t}} & {|E_t|^5} \\
\end{array}
$$

\captionof{table}{Sub-orbit breakdown for graphlets up to order 4. Edges that belong to symmetric sets are always lexically sorted (smaller plex count links first, lexical sort on same size links) before assigning them a specific sub-orbit instance.
For a set with $n$ elements, the number of $k$-combination with repetitions is denoted  by
$\left({{{n}\choose {k}}}\right)$ 
where $\left({{{n}\choose {k}}}\right) = {{{n+k-1}\choose {k}}}$. } 
\label{tab:explain}

\end{center}


\begin{figure}
\centering
\includegraphics[width=\linewidth]{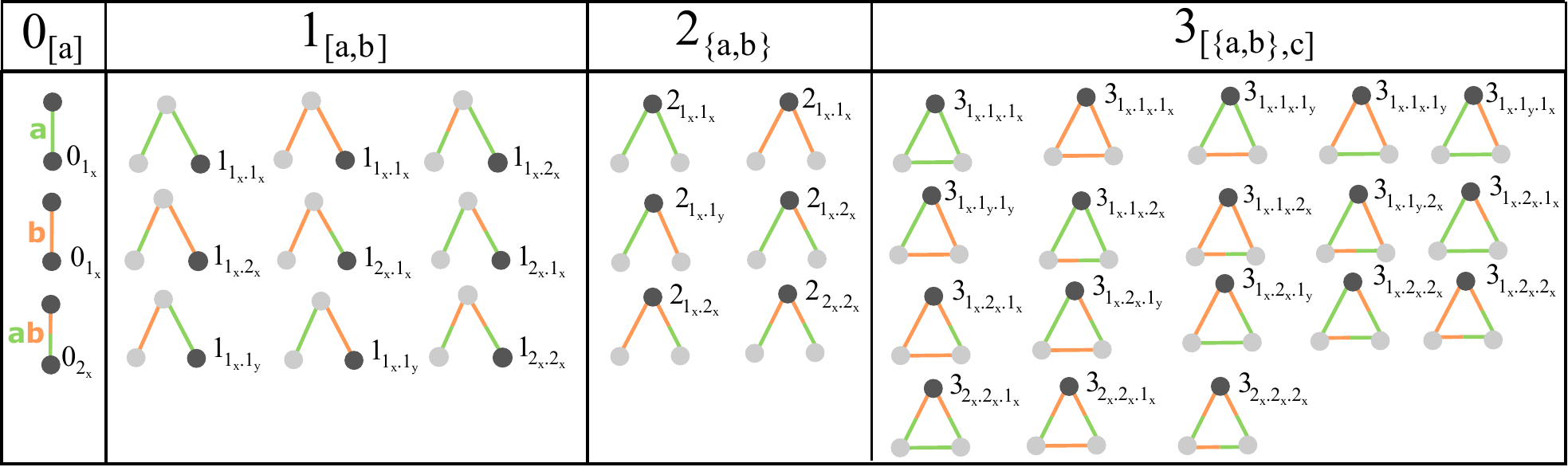} 
\caption{Reducing the number of sub-orbits: plex-count multiplex with distinct links inside orbits. The first 4 orbits and 36 sub-orbits of  a 2-plex network with $ {\cal R} = \{a, b\}$. Each plex is identified by a different color.
Each of the nodes with dark gray color is identified with the specific second reduction sub-orbit.}
\label{fig:separated}
\end{figure}

\subsection*{Reducing the number of sub-orbits}

A multiplex graph is a special type of labeled graph in which each edge is labeled only with a single label - an element of the set $E_t$. 
The number of sub-orbits grows exponentially with the linear increase in the number of plexes in the network. Even for the simplest orbit 0 -- the degree -- the number of sub-orbits grows as $2^d$.  This limits the application of graphlets for real data even for networks with small number of plexes.  In order to address this problem, we propose two  different ways for reducing the number of sub-orbits.

Let us first consider an example for which $ {\cal R} = \{a,b,c \}$, so that an edge of the multiplex is represented by an element of the set $E_t = \{ a,b,c ,ab,ac, cb,abc \}$. In this case, there are 49 wedge-path sub-orbits.  We write  $e_1.e_2$ for a wedge-path sub-orbit, where $e_1, e_2 \in E_t$. Thus, for example, $ab.abc$ denotes a sub-orbit with relation types $a$ and $b$ associated with the first edge and $a,b$ and $c$ with the second edge. The set $O_{wp}$ then contains 49 wedge-path sub-orbits $O_{wp}=\{a.a, b.b, c.c , \cdots, abc.ab, abc.ac, abc.bc, abc.abc \} $. The set of triangle sub-orbits is similarly large $|O_{tri}|=196$.

For a given orbit, the size of the set of sub-orbits (corresponding to this orbit) grows exponentially with the linear increase in the number of plexes in the networks (as shown in the Table \ref{tab:explain}). In fact, the number of sub-orbits of a given orbit is polynomial to $|E_t| = 2^d -1 $. Therefore, in order to reduce the size of the set of sub-orbits, one needs to reduce the set $E_t$.   


The simplest possible reduction 
of the $E_t$ is to the set $E_t^{'}$ defined as  $E_t^{'}=\{ |e| \hspace{0.1cm} | \hspace{0.1cm} e \in E_t\}$. Thus, in the above example, the set  $E_t = \{ a,b,c ,ab,ac, cb,abc \}$ reduces to the set $E_t^{'}=\{1,2,3 \}$. In another words, an edge of a multiplex, which has been defined as an element of the set $E_t$ representing different relation types, is now (after reduction) defined as an element of the set $E_t^{'}$ representing the `strength' (plex count) of the original edge. This reduction leaves us with $d$ possible edge types.    The reduced multiplex is called plex-count  multiplex. 



For the above example, the reduced wedge-path sub-orbit set is now 
$O_{wp}^{'}=\{1.1,1.2,1.3,2.1,2.2,2.3,3.1,3.2,3.3 \}$, with $|O_{wp}^{'}|=9$, while the triangle sub-orbit set has size $|O_{tri}^{'}|=18$. Furthermore, every orbit set has a cardinality that is $poly(d)$ instead of $poly(2^d-1)$. 
In general, the orbit set is now defined as $O_i^{'}=\{ e_1^{'}.e_2^{'}.e_3^{'}..e_k^{'} \hspace{0.1cm}| \hspace{0.1cm} e_j^{'} \in E_t^{'}\}$ where $e_i^{'}=|e_i|$.

The reduction of $E_t$ equalizes every $e_i$ that has the same plex count. This greatly reduces the information in the transformed graph. To maintain some of the original information, we can separate two different links \textbf{inside} a particular orbit that have the same plex count. Suppose we have two links $e_k=\mathbf{ab}, \hspace{0.1cm}e_v=\mathbf{bc}$, that belong to the following triangle edge list $\mathbf{ab}.\mathbf{bc}.abc \in O_{tri}$. Clearly $ |e_k|=|e_v|=2$ however $e_k\neq e_v$.  With the first reduction both of these links are transformed to $2$, so that $ab.bc.abc \to 2.2.3$. 
However, in order to retain the fact that  $e_k$ and $e_v$ are  different, we introduce a new reduction rule $e_i^{''}=d \times I_i+e_u^{'}$, where  ${I}_i$ represents a distinct id of the original link $e_i$. We assign a different ${I}_i$ for every distinct link that shares the same plex count with another link \textbf{inside} the orbit.
With this mapping, the triangle $ab.bc.abc$ corresponds to $ 2.5.3$, where $e_1^{''}=3 \times 0+2=2$ and $e_2^{''}=3 \times 1+2=5$, since $d=3$ and   
$e_u^{'}=|ab|=|bc| =2$. 
Specifically for a given orbit list $e_1.e_2.\cdots.e_l$, the distinct links (of the same plex size) are given different indices from left to right, starting with 0. 
This reduction of the set $E_t$ adds some new labels in the set $E_t^{'}$ resulting in the set $E_t^{''}$. For example, the set of sub-orbits to the wedge-path orbit is now:  $O_{wp}^{''}=\{1.1,\mathbf{1.4},1.2,1.3,2.1,2.2,\mathbf{2.5},2.3,3.3\}$. Although the size of the set  $O_{wp}^{''}$
is increased compared with the set $O_i^{'}$, the cardinality of the set $|O_i^{''}|$ is still polynomial to $d$. We have $|O_{wp}^{''}|=11$, and $|O_{tri}^{''}|=34$ for $d>2$ and so on. 
This reduction results in a multiplex which will be called plex-count multiplex with distinct links inside orbits. 
Figure \ref{fig:separated} shows all possible sub-orbits of a plex-count multiplex with distinct links inside orbits with two plexes. The exact sizes of the orbit sets, for both reductions, are shown in the tables \ref{tab:orbs_nonr}  and \ref{tab:orbs_nonr_om}.

\begin{table}
\begin{center}
$$
\begin{array}[H]{llll}
\mbox{Orbit} & \mbox{Orbit Alias} & \mbox{Orbit class} & \mbox{Size of orbit} \\ \hline \\
0 & \mbox{Degree} & {0_{[a_i]}} & {d} \\
1 & \mbox{Wedge Path} & {1_{[a_i,a_j]}} & {d^2} \\
2 & \mbox{Wedge Star} & {2_{\{a_i,a_j\}}} & {( {{d}\choose{2}}) } \\
3 & \mbox{Triangle} & {3_{[\{a_i,a_j\},a_k]}} & { ( {{d}\choose{2}})*d } 
\end{array}
$$
\captionof{table}{Reducing the number of sub-orbits: the size of the orbits 0,1,2, and 3 for a plex-count multiplex network.   $a_i,a_j,a_k \in E_t^{'}$ represent arbitrary plex-count link. 
}
\label{tab:orbs_nonr}
\end{center}
\end{table}

\begin{table}
\begin{center}
$$
\begin{array}[H]{llll}
\mbox{Orbit} & \mbox{Orbit Alias} & \mbox{Orbit class} & \mbox{Size of orbit} \\ \hline \\
0 & \mbox{Degree} & {0_{[a_x]} } & {d} \\
1 & \mbox{Wedge Path} & {1_{[a_x,b_y]} } & {d^2+(d-1)} \\
2 & \mbox{Wedge Star} & {2_{\{a_x,b_y\}} } & { ({{d}\choose{2}  }) + (d-1)} \\
3 & \mbox{Triangle} & {3_{[\{a_x,b_y\},c_z]}} & { \left[  ({{d}\choose{2}  }) + (d-1) \right]*d+[(d+1)*(d-1)]} \\
3* & \mbox{Triangle for $d>2$} & {3_{[\{a_x,b_y\},c_z]}} & { \left[  ({{d}\choose{2}  }) + (d-1) \right]*d+[(d+1)*(d-1)] +(d-1)} 
\end{array}
$$
\captionof{table}{Reducing the number of sub-orbits: the size of the orbits 0, 1, 2, and 3 for a plex-count multiplex with distinct links inside orbits.  $a,b,c \in E_t^{'}$ represent arbitrary ($a$, $b$ or $c$)-plex count links while $x,y,z$ are indexes that distinctly identify links of the same plex size: for two links $a_x$ and $b_y$, if $a=b$ then $x=y$. }
\label{tab:orbs_nonr_om}
\end{center}
\end{table}

\subsection*{Graphlet metrics}

Here several graphlet metrics for a multiplex network are defined. 
Given a node, its \textit{graphlet degree of a sub-orbit} is the number of times the node is touched by the sub-orbit.  The size of a graphlet is the cardinality of its node set. 
Let $G=(V, E^1, \ldots, E^D)$ be a multiplex network. We define a  $(n,k)-$\textit{signature vector} of the node $i$, $SI_i(n,k)$, as a vector of graphlet degrees of the node’s (lexicographically ordered) sub-orbits, for all graphlets up to the size $n$ of the $k-$plex  $G=(V, E^{\alpha_1}, \ldots, E^{\alpha_k})$.    
%


Let $G$ be a multiplex network with $N$ nodes and $D$ plexes. For large $D$, computing the signature vector of a node is impossible and we restrict ourselves, in this case, to sub-multiplexes consisting of $k$ plexes.   We first define a $k$-plex and then construct the vertex $(n,k)-$signature vector of the $k-$plex for graphlets up to size $n$.  In this way, for each vertex and a given $k-$plex $G(V,E^{\alpha_1}, \ldots, E^{\alpha_k})$, we obtain its signature vector of length $|SI(n,k)|$. Next, we construct an $N\times |SI(n,k)|$ matrix whose rows are the $(n,k)-$signature vectors for each vertex. For a given multiplex network $G$ and its sub-network with plexes $\alpha_1, \ldots, \alpha_k$, we compute Spearman's correlation coefficients between all pairs of columns of the above described matrix and present them in a $|SI(n,k)| \times |SI(n,k)|$ symmetric matrix which is termed \textit{graphlet correlation matrix} of the $k-$plex network $G(V, E^{\alpha_1}, \ldots, E^{\alpha_k})$. In this way, the network topology and its local direction patterns, regardless of network size (the number of vertices) and network volume (the number of edges), are summarized into a $|SI(n,k)| \times |SI(n,k)|$ matrix for the $k-$plex $G(V, E^{\alpha_1}, \ldots, E^{\alpha_k})$. 
Since the Graphlet Correlation Matrix of a network $G_1$ is such that devises a network statistic based on correlations between the node properties across the multiplex orbits and sub-orbits, we can examine the network topology of two networks by introducing a Graphlet Correlation Distance (GCD). Moreover, for two graphs, $G_1$ and $G_2$ and their graphlet correlation matrices $GCM_{G_1}$ and $GCM_{G_2}$, which are clean of redundancies and encode the information about the local network topology of a multiplex network that we examine, the \textit{graphlet correlation distance} is defined as the Euclidean distance of their upper triangle values:
\begin{equation}
GCD(G_1,G_2)=\sqrt{\sum_{i=1}^d \sum_{j=i+1}^d (GCM_{G_1}(i,j)-GCM_{G_2}(i,j))^2}
\end{equation}

This metric has been used for a single-plex network by Pr\v{z}ulj et al. \cite{yaverouglu2014revealing}. Here the same distance has been adopted for $k$-plex network, $k\geq 2$, representing the network topology through the local connectivity.



\subsection*{Data}

\noindent
\textit{Synthetic data} -- We generate 2000 synthetic multiplex networks, each having two plexes which are generated from the same algorithm, using four graph algorithms: Erdos-Renyi (ER), Watts-Strogatz (WS), Barabashi-Albert (BA) and a modified BA algorithm (PL- powerlaw cluster)\cite{holme2002growing}. The size of the network varies between $N=100,200,300,400,500$ and both plexes are generated using the same algorithm with the same set of parameters. For the PL algorithm the triangle forming probability is 0.8, while WS has a rewiring probability of 0.01. The other parameters are either the size of the connected component or the sparsity probability. These are derived as follows: $p \in \{0.2,0.35,0.5,0.75,0.8\}, k=\floor{p*N}$, where p is the sparsity probability for ER, while k represents the initial connected component parameter for BA and PL, and the size of the initial neighbours in WS.
This is fully explained in the SI.

\noindent
\textit{Economic Trade Networks Data} -- International trade network data from the most recent available year (2000), provided in~\cite{feenstra2000nber}, are used to construct a ``Multiplex International Trade Network’’, in which plexes represent products, nodes are countries and a link between country $i$ and $j$ in the plex $\alpha$ exists if at least one of the countries is a significant exporter of the product $\alpha$ to the other country. The full explanation of how the network is created is given in the SI. The final network contains of $N=125$ countries (nodes) and $D=957$ products (plexes). From this network we then focus on a subset of products, or combine products into product categories by their hierarchical Standard International Trade Classification (SITC) code. 

\noindent
\textit{Social Networks Data} -- 
Data collected from 75 villages in the region of Karnatka \cite{banerjee2013diffusion} was used. Each node represents an individual with age ranging from 18 to 57. The individuals were asked how they interact between each other in the village, across 12 different aspects of the everyday life: visiting other's homes, who they were inviting to visit their's home, kin, nonrelatives with whom they socialize, from who they receive medical advice, those from whom would borrow money and from whom they would lend money, those from which they would borrow material goods (kerosene, rice, etc.), those to whom they would lend material goods, giving or getting advice, people with whom they go to pray (at temple, church or mosque). The data is organized as a multiplex network with a number of nodes dependent on the village size and $D=12$ plexes.

\section*{Results and Discussions}  
\label{sec:results}

A typical multiplex network has large number of plexes and, therefore, a full graphlet analysis is computationally infeasible.  Moreover, a network might contain plexes, which are less significant or contain a small number of links. Because of this, often one might want to focus on a smaller number of plexes to be analyzed. Two different strategies have been adopted to address this problem. First, when the problem in question is such that full graphlet analysis is needed, we consider for a given multiplex network with $d$ plexes, the set (or well-defined subset) of all $k-$plex networks $G(V, E^{\alpha_1}, \ldots, E^{\alpha_k})$ such that $\alpha_1 \neq \alpha_2 \neq \ldots \alpha_k$,  where $k$ is a small number, typically $k=2,3,4$.   Second, when the problem to be addressed allows sub-orbit count reduction, we consider both reduced multiplex constructions described in the previous sections, namely plex-count multiplex and plex-count multiplex with distinct links inside orbits.  In order to compare results for computing graphlets using full multiplex and two reduced multiplexes, graphlet correlation matrices (GCMs) are computed  for all $k-$plex networks. From these GCMs we then analyze only those sub-orbit pairs for which significant correlations (above 0.7) exist in more than 60\% of these $k-$plex networks.

Results using all three approaches (full multiplex and two reduced multiplexes) are comparable and similar to each other for two data sets from two different domains: economic world trade networks and social networks, analyzed in more details in this section.  For this reason, we only present and discuss results when the second orbit size reduction is employed. For better clarity orbits are represented as $O_{i_x.j_y.k_z}$, where $O$ is the orbit number, $i,j,k$ are the plex counts and $x,y,z$ are the distinction indices (different if the original plex sets are different). The links are ordered by the orbit class definition on Table \ref{tab:explain} (edges inside symmetric sets being ordered lexically). As an example the sub-orbit $3_{13.4.25}$ is rewritten as $3_{2_x.1_x.2_y}$.

\subsubsection*{Economic Trade Networks}














The products in the economic trade network are labeled with a hierarchical 4 letter code (SITC Code). The first letters of the code refer to more general product classes such as products of animal or mineral origin, while the full code refers to more specific products such as skimmed milk or pork. The full network contains info for 957 specific products, which are arranged as a multiplex (957-plex) network. 

In order to find more general conclusions we focus on 2 letter product class pairs. From each of these product classes we then uniformly chose up to 2000 individual products and create 2-plex networks. For example, if the chosen product classes are paper and furniture products, then from each of them we can create 2-plex networks using (craft paper, leather furniture) or (newspaper rolls, wood furniture) etc. 


\begin{figure}[h!] 
\centering
\includegraphics[width=\textwidth]{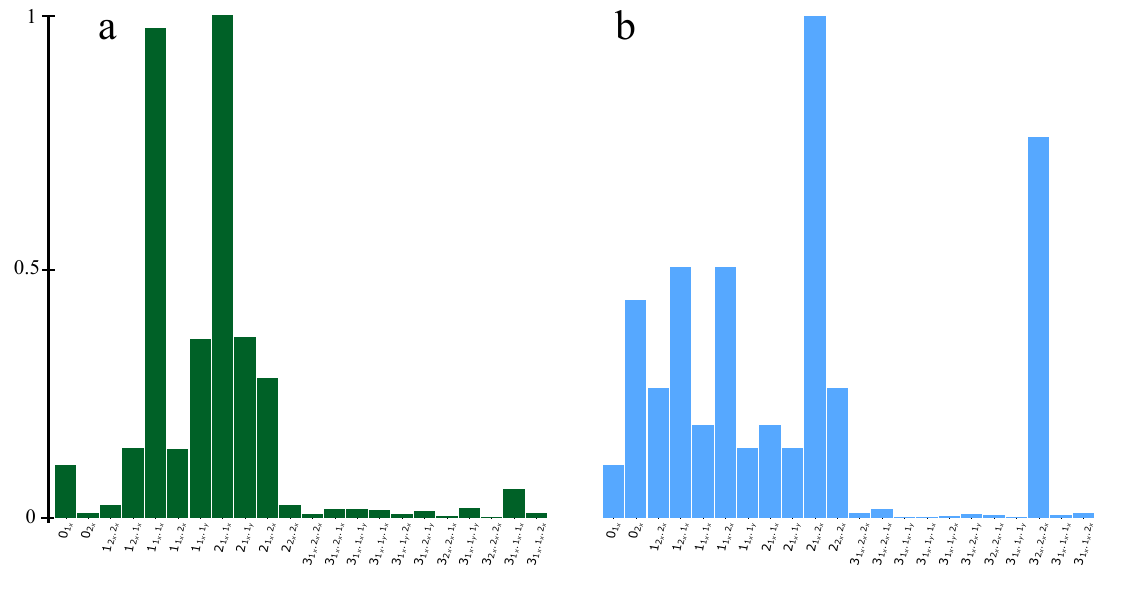}
\caption{ Histograms containing normalized frequencies of 22 (second reduction) sub-orbits from 2-plex economic trade networks (a) and social networks (b). 4950 2-plex social networks and 15390 2-plex economic trade networks were analyzed. Detailed explanation how the 2-plex networks were constructed is provided in the main text and in the SI.} 
\label{fig:histogramsom}
\end{figure}

Figure \ref{fig:histogramsom}a shows the histogram (normalized frequencies) of the sub-orbits in the economic network. Dominant graphlets  are wedges, reflecting how economic networks are built: as trading networks between two countries. Moreover, in economic trade networks wedge paths and wedge stars are (almost) equally represented.      
We now examine the correlations between graphlet sub-orbits. 
We randomly chose 100 product class pairs. For each class pair, we select up to 2000 2-plex networks constructed from individual products belonging to the two classes respectively. For every product 2-plex network, we compute its GCM and remember the significant correlations ($>0.7$). Subsequently for each class pair we retain strong correlations appearing in more than 60\% of the individual networks. To generalize about economic networks we than examine correlations appearing in more than 90 \% of the 100 random product class pairs. 

\begin{figure}[h!]
\centering
\includegraphics[width=\textwidth]{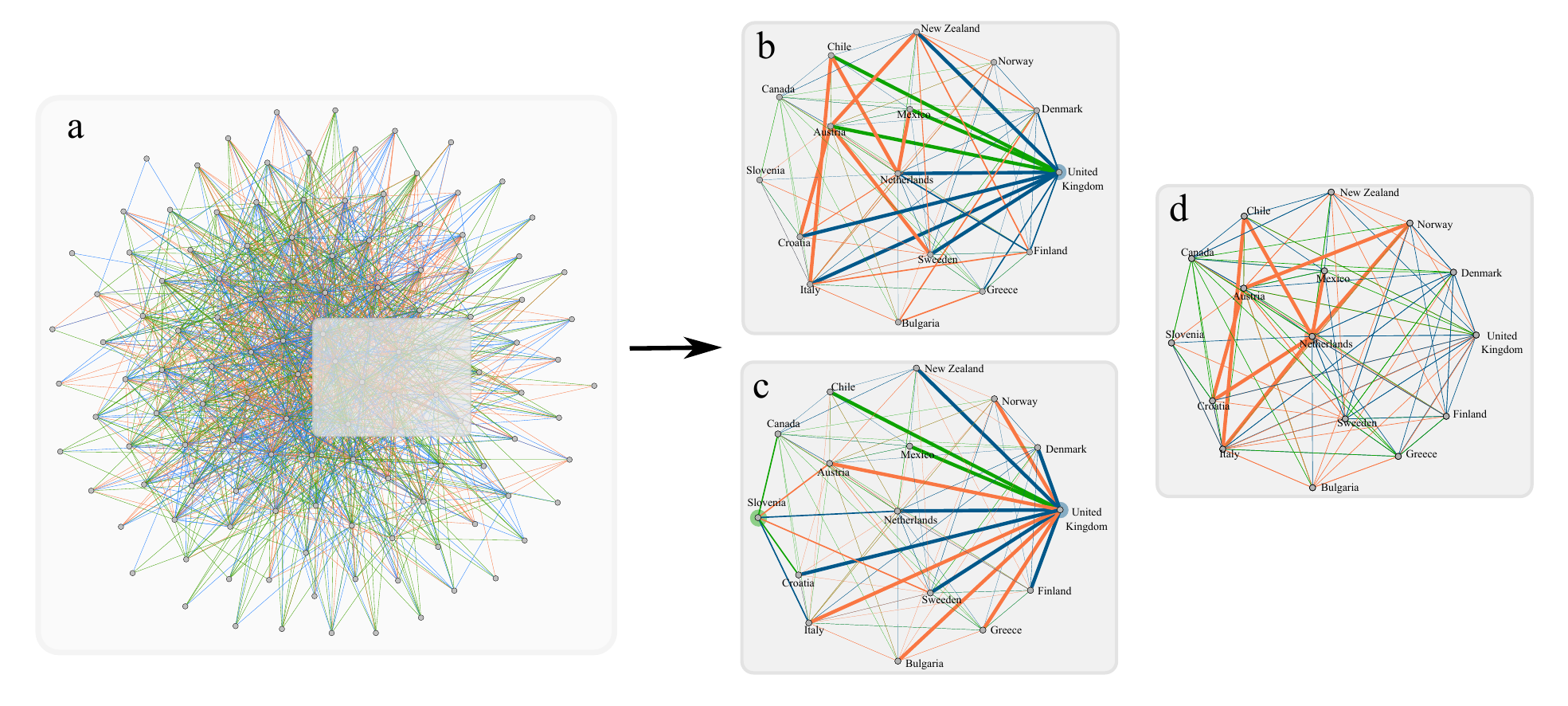}
\caption{ (a) Full international trade network of 2 plexes: diary and meat products. The green and orange links represent trade of dairy and meat respectively, while blue links represent both dairy and meat trade. Small induced subgraph from United Kingdom's trade network is further visualized for better understanding for highlighting different structural patterns: (b) strong preferences of trading with neighbors of node's traders, (c) trade hubs of the network, (d) triangle single product relations. }
\label{fig:econ_cors}
\end{figure}

The significant correlations that exist in a majority of these pairs provide a more general overview of graphlet correlations in economic trade networks. The full correlation tables are presented in the SI. Part of the correlations that emerged and are more interesting and nontrivial to our understanding are shown on Figure \ref{fig:econ_cors} and described thoroughly below. For these networks, only positive correlations were found, which in the case of wedges and triangles follow an interesting pattern of behavior which also emerged when we employed the same method, for 3 and 4-plex networks. This can be reviewed in the figures provided in the SI, since here we describe in details only the 2-plex correlations and their meaning and importance.

The correlation  $0_{2_x}$ and $2_{2_x.1_x}$ appears in all 100 randomly chosen economic 2-plexes. This implies that there are trade relations from a certain node, some of which are strong (trading both products), while others are weak (trading one product). Another similar strong correlation is between $2_{2_x.2_x}$ and $2_{1_x.2_x}$ (appearing in 98 \% of observed networks).  From these correlations we can infer that certain countries(nodes) act as a type of network hub, `specializing' in two specific products (links), which they trade interchangeably with different countries. Furthermore from the frequency histogram on the same networks, it is clear that strong relations are not common (which makes sense as we choose random two products), however wedge star relations of the type $2_{2_x.1_x}$ appear more frequently. 

We also observe triangle trade relations in the economic networks. The sub-orbits  $0_{1_x}$ and $3_{1_x.1_x.1_x}$ are strongly correlated in 96 \% of observed networks. This implies that countries form trade deals with a specific product. This is supported also by the histogram data, which shows that single product triangles appear more frequently than other triangles in the networks.
Furthermore we observe correlations between  $0_{2_x}$ and $3_{2_x.1_x.1_y]}$ or  $3_{2_x.1_x.1_x}$, which appear in 97\% of sampled networks. Therefore, when a country trades one or two products with two different countries, these countries also form trade deals with each other (with a single product), meaning that one of the countries in the triangle is a 'stronger' trader, trading in two products, while the other two countries mostly trade with a single product. This construction is furthermore supported with the correlation between  $3_{2_x.1_x.1_x}$ and  $3_{2_x.1_x.1_y}$, appearing in 95 \% of networks, suggesting that there are 'trade triangles' that are composed of a strong trade link $2_x$, (which are rarer) and single product trade links $1_i$ composed of any of the two products.
Moreover, the formation of these deals can be supported with the following correlations: $2_{2_x.1_x}$ and  $3_{2_x.1_x.1_x}$, with 93 \% of observed networks; $2_{2_x.1_x}$ and  $3_{2_x.1_x.1_y}$, with 93 \% of observed networks; and $2_{1_x.1_x}$ and  $3_{1_x.1_x.1_x}$, with 92 \% of observed networks.
This implies that some of the wedge trade relations are  closed into triangles with a single product trade link.
The triangle relations are represented in Fig. \ref{fig:econ_cors} b, where we can see that United Kingdom acts as a hub for trading meat and dairy (thus having a number of trade relations), and it forms triangle $3_{1_x.2_x.1_x}$ with Austria and New Zealand, or in Fig. \ref{fig:econ_cors} d we have the triangle $3_{1_x.1_x.1_x}$ between Chile, Italy and the trading hub Netherlands.

The concept of economic complexity has been recently introduced \cite{hidalgo2007product,hidalgo2009building,tacchella2012new,cristelli2013measuring,stojkoski2016impact} with the aim to reflect the amount of knowledge that is embedded in the productive structure of an economy. Capability-driven economic competitiveness has been analyzed using three methods: methods for reflections, fitness-complexity method, and modified fitness-complexity method. Two simple measures have been introduced both related to degrees: the first is country degree (in the bipartite network) and is called diversity and the second is the product degree (in the bipartite network) and is called ubiquity. Here graphlet analysis provides another view of the economic competitiveness. For those countries with large diversity (computed as in \cite{hidalgo2007product}, for example) we found that sub-orbits  $2_{2_x.2_x}$ and $2_{1_x.2_x}$ are correlated as well as sub-orbits   $1_{2_x.2_x}$ and $1_{1_x.2_x}$   but not $2_{2_x.2_x}$ and $2_{1_x.1_x}$ or other pairs for wedges. This implies that the countries with large diversity are structurally (locally) well described with wedges and have both double and single plex links. On the other hand, the fact that correlations between degrees and triangles are found among sub-orbits $0_{2_x}$ and $3_{2_x.1_x.1_y}$ or $0_{2_x}$ with $3_{2_x.1_x.1_x}$,  $0_{1_x}$ and $3_{1_x.1_x.1_x}$, and not between other degree-triangle pairs, in particular not  between $0_{2_x}$ and $3_{2_x.2_x.2_x}$ provide evidence that that the countries with small diversity tend to form correlated triangles. 


\subsubsection*{Social Networks}

Social network data is organized as 75 (the number of villages) multiplex networks with a number of nodes dependent on the village size and $D = 12$ plexes. Each 12-plex network is further separated into all different combinations of 2-plexes and the graphlet analysis, similar to one performed for economic data, is carried out for social data as well. 
We run our analysis for all possible relation pairs, 66 in total. For each relation pair we create 2-plex networks for each village, then we extract the strong correlations that appear in a majority of villages $(>60\%)$. These correlations are assumed to be representative of the specific relation pair. By finding correlations that appear in a majority of relation pairs $(>80\%)$ we aim to find more general correlations that appear in social networks. 

 \begin{figure}[h!]
\centering
\includegraphics[width=\textwidth]{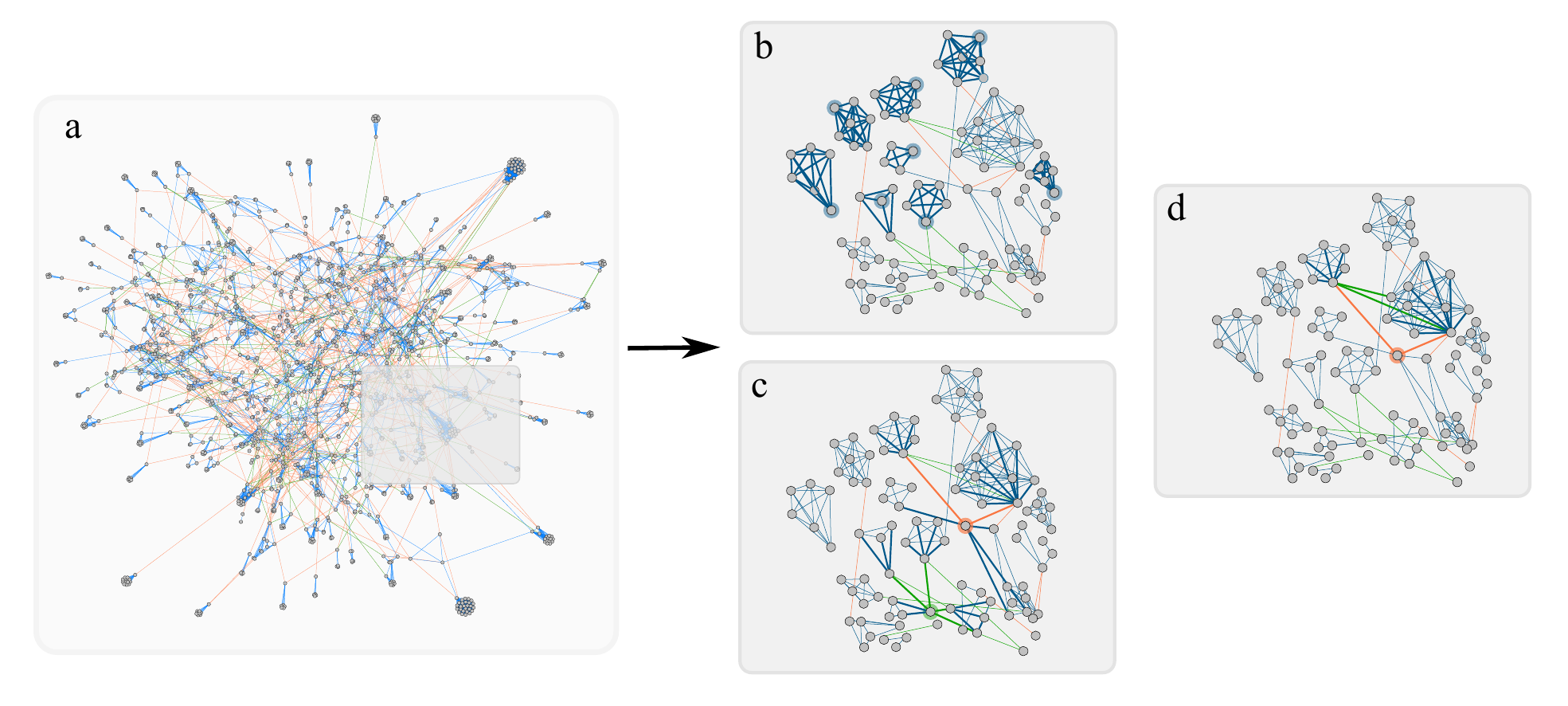}
\caption{  (a) 2-plex social network from one of the villages (village 28): give advice (orange links) and friends relations (green links). Blue links represent both, giving advice and friend relations. Focus, again, for better visualization is given for a smaller induced subgraph. (b) Highlights the strong tie cliques in the network; (c) structural holes of the network, and (d) links between the structural holes and the outreach node of a clique. }
\label{fig:soc_cors}
\end{figure}

Figure \ref{fig:histogramsom}b depicts the histogram (normalized frequencies) of the sub-orbits in the social network. Dominant graphlets are sub-orbits: degree $0_{2_x}$, wedges with at least one $2_x$ link, that is $1_{1_x.2_x}$, $1_{2_x.1_x}$, $1_{2_x.2_x}$, and $2_{1_x.2_x}$, $2_{2_x.2_x}$, and the triangle $3_{2_x.2_x.2_x}$). 
A significant correlation (100\% of all social relation combinations, in the majority of villages) was found between degree sub-orbit $0_{2_x}$ and triangle sub-orbit $3_{2_x.2_x.2_x}$ (shown in Fig. \ref{fig:soc_cors}b). 
In the seminal work Granovetter \cite{Granovetter1973} suggested the strength of dyadic ties to be the tool linking micro and macro levels of sociological theory. He showed that dyadic ties are related to larger structures by implementing the following principle: the stronger the tie between two individuals, the larger the proportion of individuals to whom they will \textit{both} be tied. The impact of this principle on diffusion of influence and information, mobility opportunity, and community organization is well documented. This principle has been supported by providing evidence that the triads in which two ties are strong and the third is absent are unlikely to occur.    
Following Granovetter here we suggest multiplexity as a way of indicating a strong tie. Thus, the tie $2_x$ (in which both plexes are presented) is called strong tie, while ties $1_i$ are weak ties.  We also call $3_{2_x.2_x.2_x}$ a strong triangle. The graphlet analysis shows that triangles other than $3_{2_x.2_x.2_x}$ are unlikely to occur;  moreover, the occurrence of the strong triangles is highly correlated with the occurrence of the strong ties, supporting Granovetter’s principle. However, in our case, this support is direct. The occurrence of wedges $1_{2_x.2_x}$ and $2_{2_x.2_x}$ are significant (in contrast to \cite{Granovetter1973} in which such wedges are unlikely to occur).  Moreover, we found significant correlation in 89\% of 2-plexes between triads in which two ties are strong and the third is absent. In another words, we found that the wedge stars $2_{2_x.2_x}$ and the wedge paths $1_{2_x.2_x}$ are strongly correlated.    
 
Another concept related to weak ties is the concept of structural holes \cite{holes1992social}, introduced to explain the origin of differences in social capital. An individual holds certain positional advantages/disadvantages from how she/he is embedded in neighborhoods or other social structures. A structural hole represents the gap between two individuals who have complementary sources of information. A simple measure of structural holes in a network is the bridge count. According to Granovetter, no strong tie is a bridge \cite{Granovetter1973}.   Several significant graphlet correlations support the concept of structural holes. Thus, significant correlations ($>0.7$) are found in 100\% of the tested 2-plex networks between wedge star $2_{1_x.2_x}$ with $1_{1_x.2_x}$,  also $1_{1_x.1_x}$ with $1_{1_x.2_x}$ in 95\% of networks.  When looking at the data we found that these structures appear often around the same person. Furthermore we observe the following correlations between $0_{1_x}$ and these structures:  $2_{1_x.2_x}$, $1_{1_x.2_x}$ with 100\% of relations pairs, also  $1_{1_x.1_x}$ with 97\%, and $2_{1_x.1_x}$ with 94\% of relation pairs. 
If we examine the connections that these people have within the network, we can conclude that they behave as structural holes, meaning they connect different social cliques (households). This can be viewed on Figure \ref{fig:soc_cors}c which depicts several examples of individuals that are mediators between two or more households as part of different cliques. A correlation that appears in 83\% of networks is between wedge paths $1_{1_x.1_y}$ with $1_{1_x.2_x}$. This correlation again signifies that the network has individuals that serve as a connection between two cliques. Furthermore, these graphlet structures can also detect individuals inside the clique that are linked to outside, having broader communication reach. This can be observed on Figure \ref{fig:soc_cors}d where only a few people inside a clique are connected to an outside person. For the full list of correlations we refer the reader to the SI. 

We remark that these conclusions are based on the analysis of a single data set consisting of 75 social networks. Therefore, more thoughtful social network analysis for reaching more general conclusion is needed, which is, however, beyond the scope of this paper and will be provided in a future study.

\begin{figure}[h!] 
\centering
\includegraphics[width=\textwidth]{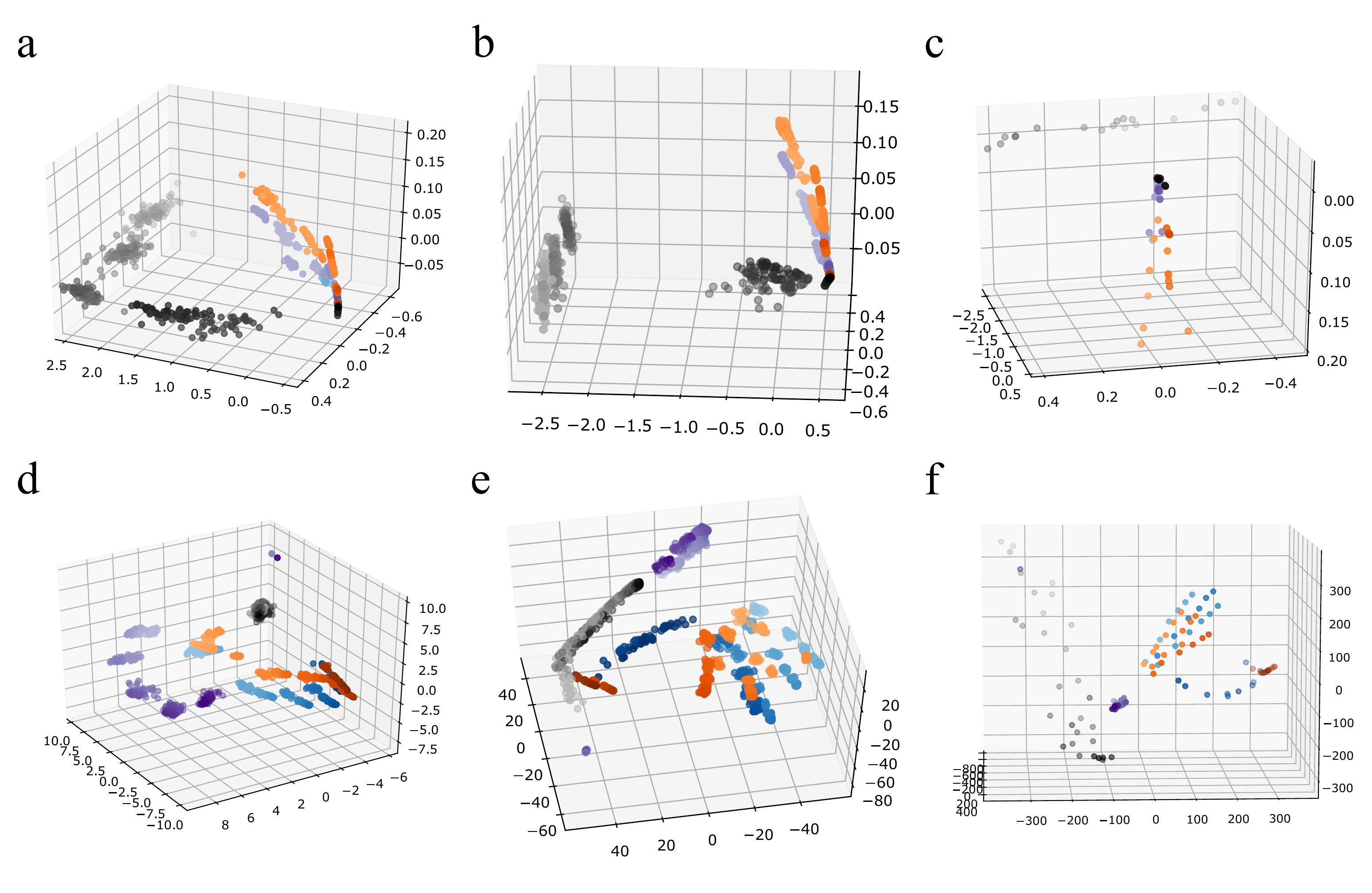}
\caption{ Synthetic networks: 3D Visualization using graphlet correlations. Purple points: Erdos-Renyi (ER) network, gray points: Watts-Strogatz (WS) network, blue points: Barabashi-Albert (BA) network, and orange points: modified BA (PL- power-law cluster) network. (a), (b), and (c) flattened networks of 2, 3 and 4 plexes; (d), (e), and (f) 2-plex, 3-plex and 4-plex networks, respectively. }\label{fig:3d_synth}
\end{figure}

\begin{figure}[h!] 
\centering

\includegraphics[width=\textwidth]{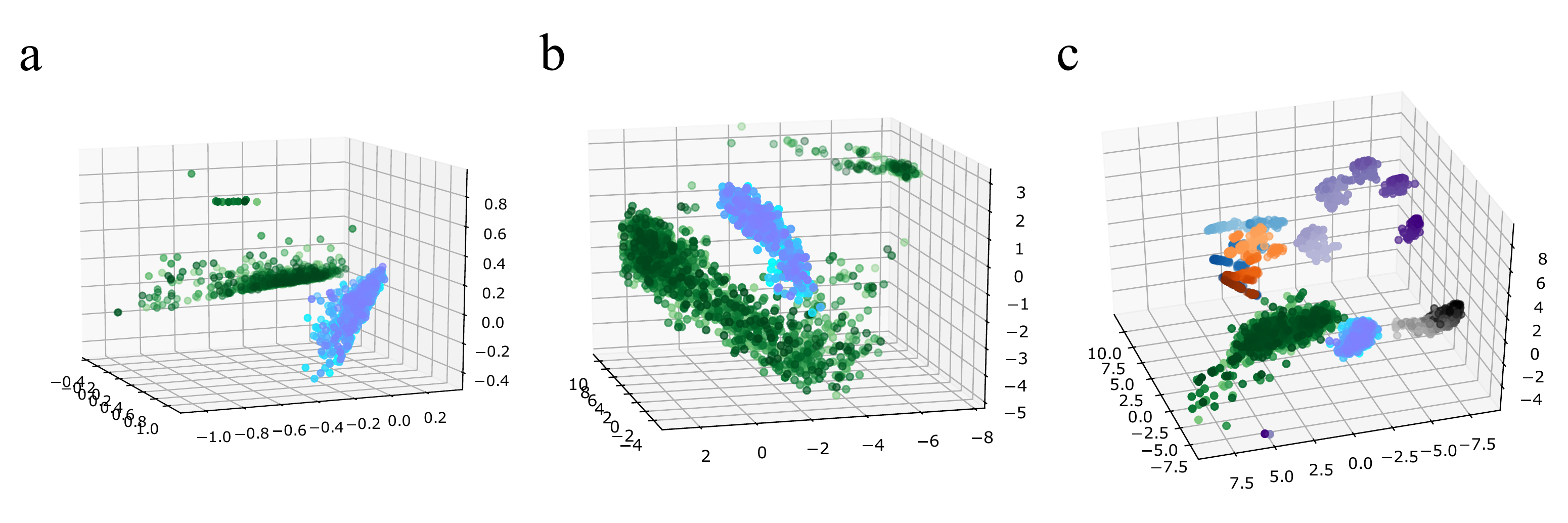}
\caption{ 3D visualization using graphlet correlations. Green points are the economic 2-plex networks while social 2-plexes are shown with cyan points. (a) For better visualization, 2\% of 2-plex flattened networks are shown; (b) full 2-plex networks; (c) social and economic 2-plex networks positioned together with the synthetic 2-plex networks. } \label{fig:3d_econ}
\end{figure}

\section*{Conclusions}

Graphlets are a powerful tool for analyzing local network structure. Multiplex networks, multilayer networks, and networks with node and/or link (categorical) attributes are pervasive and graphlet analysis developed here can further enhance our understanding of complex networks. Graphlets provide discriminatory property for different type networks. Even a simple graphlet histogram plot of economic and social networks, see Fig.~\ref{fig:histogramsom}, shows the differences between these structures and provides evidences on how the networks are built.  Wedges occur more often in economic networks rather than social networks, indicating the tendency of a country to produce/trade of a product in local structure of triads which are not closed (that is, wedges not triangles).   Wedges (open triads) also appear in the social networks, however the dominant graphlets in social networks are triangles (closed triads). If the multiplex is the indicator of the strong tie, the graphlet analysis provides another evidence for the concepts of strong/weak ties and structural holes. In contrast to the work of Granovetter\cite{Granovetter1973}, however, in our work related to a single data set consisting of 75 social networks, wedges with only strong ties are not only present but they are strongly correlated.

Graphlets can also provide clustering.  Graphlet correlation matrix of a given network is represented as point in some multidimensional space, which is then visualized in 3D space, by using multidimensional scaling. This has been demonstrated with synthetic networks in Fig. \ref{fig:3d_synth}. Several conclusions can be drawn from the figure: (1) different graphs are clearly separated, (2) same graphs with different sparsity are also distinguishable, and (3) aggregating a multiplex network in a single-plex network provides less information on the network structure.  Graphlet correlation matrices for economic and social networks are visualized in Figure \ref{fig:3d_econ}. We again notice a good separation of the economic and social networks. This is present both in the flattened and the 2-plex representation. However, due to the additional plex information, 2-plex graph clouds have better resolution. The final part of Figure \ref{fig:3d_econ} shows the synthetic and real networks in the same 3d space, and again there exists a strong separation among the different graph types. The full description on how graphlets can be used for clustering will be provided in a separate manuscript.

\bibliography{main}

\begin{thebibliography}{10}
\expandafter\ifx\csname url\endcsname\relax
  \def\url#1{\texttt{#1}}\fi
\expandafter\ifx\csname urlprefix\endcsname\relax\def\urlprefix{URL }\fi
\expandafter\ifx\csname doiprefix\endcsname\relax\def\doiprefix{DOI }\fi
\providecommand{\bibinfo}[2]{#2}
\providecommand{\eprint}[2][]{\url{#2}}

\bibitem{milo2002network}
\bibinfo{author}{Milo, R.} \emph{et~al.}
\newblock \bibinfo{title}{Network motifs: simple building blocks of complex
  networks}.
\newblock \emph{\bibinfo{journal}{Science}} \textbf{\bibinfo{volume}{298}},
  \bibinfo{pages}{824--827} (\bibinfo{year}{2002}).

\bibitem{sporns2004motifs}
\bibinfo{author}{Sporns, O.} \& \bibinfo{author}{K{\"o}tter, R.}
\newblock \bibinfo{title}{Motifs in brain networks}.
\newblock \emph{\bibinfo{journal}{PLoS Biol}} \textbf{\bibinfo{volume}{2}},
  \bibinfo{pages}{e369} (\bibinfo{year}{2004}).

\bibitem{prvzulj2004modeling}
\bibinfo{author}{Pr{\v{z}}ulj, N.}, \bibinfo{author}{Corneil, D.~G.} \&
  \bibinfo{author}{Jurisica, I.}
\newblock \bibinfo{title}{Modeling interactome: scale-free or geometric?}
\newblock \emph{\bibinfo{journal}{Bioinformatics}}
  \textbf{\bibinfo{volume}{20}}, \bibinfo{pages}{3508--3515}
  (\bibinfo{year}{2004}).

\bibitem{holland1976local}
\bibinfo{author}{Holland, P.~W.} \& \bibinfo{author}{Leinhardt, S.}
\newblock \bibinfo{title}{Local structure in social networks}.
\newblock \emph{\bibinfo{journal}{Sociological methodology}}
  \textbf{\bibinfo{volume}{7}}, \bibinfo{pages}{1--45} (\bibinfo{year}{1976}).

\bibitem{faust2010puzzle}
\bibinfo{author}{Faust, K.}
\newblock \bibinfo{title}{A puzzle concerning triads in social networks: Graph
  constraints and the triad census}.
\newblock \emph{\bibinfo{journal}{Social Networks}}
  \textbf{\bibinfo{volume}{32}}, \bibinfo{pages}{221--233}
  (\bibinfo{year}{2010}).

\bibitem{prvzulj2007biological}
\bibinfo{author}{Pr{\v{z}}ulj, N.}
\newblock \bibinfo{title}{Biological network comparison using graphlet degree
  distribution}.
\newblock \emph{\bibinfo{journal}{Bioinformatics}}
  \textbf{\bibinfo{volume}{23}}, \bibinfo{pages}{e177--e183}
  (\bibinfo{year}{2007}).

\bibitem{yaverouglu2014revealing}
\bibinfo{author}{Yavero{\u{g}}lu, {\"O}.~N.} \emph{et~al.}
\newblock \bibinfo{title}{Revealing the hidden language of complex networks}.
\newblock \emph{\bibinfo{journal}{Scientific reports}}
  \textbf{\bibinfo{volume}{4}}, \bibinfo{pages}{4547} (\bibinfo{year}{2014}).

\bibitem{sarajlic2016graphlet}
\bibinfo{author}{Sarajli{\'c}, A.}, \bibinfo{author}{Malod-Dognin, N.},
  \bibinfo{author}{Yavero{\u{g}}lu, {\"O}.~N.} \&
  \bibinfo{author}{Pr{\v{z}}ulj, N.}
\newblock \bibinfo{title}{Graphlet-based characterization of directed
  networks}.
\newblock \emph{\bibinfo{journal}{Scientific reports}}
  \textbf{\bibinfo{volume}{6}} (\bibinfo{year}{2016}).

\bibitem{trpevski2016graphlet}
\bibinfo{author}{Trpevski, I.}, \bibinfo{author}{Dimitrova, T.},
  \bibinfo{author}{Boshkovski, T.}, \bibinfo{author}{Stikov, N.} \&
  \bibinfo{author}{Kocarev, L.}
\newblock \bibinfo{title}{Graphlet characteristics in directed networks}.
\newblock \emph{\bibinfo{journal}{Scientific Reports}}
  \textbf{\bibinfo{volume}{6}} (\bibinfo{year}{2016}).

\bibitem{Boorman1976}
\bibinfo{author}{Boorman, S.~A.} \& \bibinfo{author}{White, H.~C.}
\newblock \bibinfo{title}{Social structure from multiple networks. ii. role
  structures}.
\newblock \emph{\bibinfo{journal}{American journal of sociology}}
  \bibinfo{pages}{1384--1446} (\bibinfo{year}{1976}).

\bibitem{White1976}
\bibinfo{author}{White, H.~C.}, \bibinfo{author}{Boorman, S.~A.} \&
  \bibinfo{author}{Breiger, R.~L.}
\newblock \bibinfo{title}{Social structure from multiple networks. i.
  blockmodels of roles and positions}.
\newblock \emph{\bibinfo{journal}{American journal of sociology}}
  \bibinfo{pages}{730--780} (\bibinfo{year}{1976}).

\bibitem{gould1991multiple}
\bibinfo{author}{Gould, R.~V.}
\newblock \bibinfo{title}{Multiple networks and mobilization in the paris
  commune, 1871}.
\newblock \emph{\bibinfo{journal}{American Sociological Review}}
  \bibinfo{pages}{716--729} (\bibinfo{year}{1991}).

\bibitem{Padgett1993}
\bibinfo{author}{Padgett, J.~F.} \& \bibinfo{author}{Ansell, C.~K.}
\newblock \bibinfo{title}{Robust action and the rise of the medici, 1400-1434}.
\newblock \emph{\bibinfo{journal}{American journal of sociology}}
  \bibinfo{pages}{1259--1319} (\bibinfo{year}{1993}).

\bibitem{granovetter1985economic}
\bibinfo{author}{Granovetter, M.}
\newblock \bibinfo{title}{Economic action and social structure: The problem of
  embeddedness}.
\newblock \emph{\bibinfo{journal}{American journal of sociology}}
  \textbf{\bibinfo{volume}{91}}, \bibinfo{pages}{481--510}
  (\bibinfo{year}{1985}).

\bibitem{baldassarri2007integrative}
\bibinfo{author}{Baldassarri, D.} \& \bibinfo{author}{Diani, M.}
\newblock \bibinfo{title}{The integrative power of civic networks 1}.
\newblock \emph{\bibinfo{journal}{American Journal of Sociology}}
  \textbf{\bibinfo{volume}{113}}, \bibinfo{pages}{735--780}
  (\bibinfo{year}{2007}).

\bibitem{Grossman2009}
\bibinfo{author}{Grossmann, M.} \& \bibinfo{author}{Dominguez, C.~B.}
\newblock \bibinfo{title}{Party coalitions and interest group networks}.
\newblock \emph{\bibinfo{journal}{American Politics Research}}
  \textbf{\bibinfo{volume}{37}}, \bibinfo{pages}{767--800}
  (\bibinfo{year}{2009}).

\bibitem{Maggioni2013}
\bibinfo{author}{Maggioni, M.~A.}, \bibinfo{author}{Breschi, S.} \&
  \bibinfo{author}{Panzarasa, P.}
\newblock \bibinfo{title}{Multiplexity, growth mechanisms and structural
  variety in scientific collaboration networks}.
\newblock \emph{\bibinfo{journal}{Industry and Innovation}}
  \textbf{\bibinfo{volume}{20}}, \bibinfo{pages}{185--194}
  (\bibinfo{year}{2013}).

\bibitem{Rank2010}
\bibinfo{author}{Rank, O.~N.}, \bibinfo{author}{Robins, G.~L.} \&
  \bibinfo{author}{Pattison, P.~E.}
\newblock \bibinfo{title}{Structural logic of intraorganizational networks}.
\newblock \emph{\bibinfo{journal}{Organization Science}}
  \textbf{\bibinfo{volume}{21}}, \bibinfo{pages}{745--764}
  (\bibinfo{year}{2010}).

\bibitem{Ferriani2012}
\bibinfo{author}{Ferriani, S.}, \bibinfo{author}{Fonti, F.} \&
  \bibinfo{author}{Corrado, R.}
\newblock \bibinfo{title}{The social and economic bases of network
  multiplexity: Exploring the emergence of multiplex ties}.
\newblock \emph{\bibinfo{journal}{Strategic Organization}}
  \textbf{\bibinfo{volume}{11}}, \bibinfo{pages}{7--34} (\bibinfo{year}{2013}).

\bibitem{Shipilov2012}
\bibinfo{author}{Shipilov, A.~V.} \& \bibinfo{author}{Li, S.~X.}
\newblock \bibinfo{title}{The missing link: The effect of customers on the
  formation of relationships among producers in the multiplex triads}.
\newblock \emph{\bibinfo{journal}{Organization Science}}
  \textbf{\bibinfo{volume}{23}}, \bibinfo{pages}{472--491}
  (\bibinfo{year}{2012}).

\bibitem{de2013mathematical}
\bibinfo{author}{De~Domenico, M.} \emph{et~al.}
\newblock \bibinfo{title}{Mathematical formulation of multilayer networks}.
\newblock \emph{\bibinfo{journal}{Physical Review X}}
  \textbf{\bibinfo{volume}{3}}, \bibinfo{pages}{041022} (\bibinfo{year}{2013}).

\bibitem{kivela2014multilayer}
\bibinfo{author}{Kivel{\"a}, M.} \emph{et~al.}
\newblock \bibinfo{title}{Multilayer networks}.
\newblock \emph{\bibinfo{journal}{Journal of complex networks}}
  \textbf{\bibinfo{volume}{2}}, \bibinfo{pages}{203--271}
  (\bibinfo{year}{2014}).

\bibitem{sola2013eigenvector}
\bibinfo{author}{Sol{\'a}, L.} \emph{et~al.}
\newblock \bibinfo{title}{Eigenvector centrality of nodes in multiplex
  networks}.
\newblock \emph{\bibinfo{journal}{Chaos: An Interdisciplinary Journal of
  Nonlinear Science}} \textbf{\bibinfo{volume}{23}}, \bibinfo{pages}{033131}
  (\bibinfo{year}{2013}).

\bibitem{halu2013multiplex}
\bibinfo{author}{Halu, A.}, \bibinfo{author}{Mondrag{\'o}n, R.~J.},
  \bibinfo{author}{Panzarasa, P.} \& \bibinfo{author}{Bianconi, G.}
\newblock \bibinfo{title}{Multiplex pagerank}.
\newblock \emph{\bibinfo{journal}{PloS one}} \textbf{\bibinfo{volume}{8}},
  \bibinfo{pages}{e78293} (\bibinfo{year}{2013}).

\bibitem{battiston2014structural}
\bibinfo{author}{Battiston, F.}, \bibinfo{author}{Nicosia, V.} \&
  \bibinfo{author}{Latora, V.}
\newblock \bibinfo{title}{Structural measures for multiplex networks}.
\newblock \emph{\bibinfo{journal}{Physical Review E}}
  \textbf{\bibinfo{volume}{89}}, \bibinfo{pages}{032804}
  (\bibinfo{year}{2014}).

\bibitem{de2015ranking}
\bibinfo{author}{De~Domenico, M.}, \bibinfo{author}{Sol{\'e}-Ribalta, A.},
  \bibinfo{author}{Omodei, E.}, \bibinfo{author}{G{\'o}mez, S.} \&
  \bibinfo{author}{Arenas, A.}
\newblock \bibinfo{title}{Ranking in interconnected multilayer networks reveals
  versatile nodes}.
\newblock \emph{\bibinfo{journal}{Nature communications}}
  \textbf{\bibinfo{volume}{6}} (\bibinfo{year}{2015}).

\bibitem{cozzo2015structure}
\bibinfo{author}{Cozzo, E.} \emph{et~al.}
\newblock \bibinfo{title}{Structure of triadic relations in multiplex
  networks}.
\newblock \emph{\bibinfo{journal}{New Journal of Physics}}
  \textbf{\bibinfo{volume}{17}}, \bibinfo{pages}{073029}
  (\bibinfo{year}{2015}).

\bibitem{bianconi2013statistical}
\bibinfo{author}{Bianconi, G.}
\newblock \bibinfo{title}{Statistical mechanics of multiplex networks: Entropy
  and overlap}.
\newblock \emph{\bibinfo{journal}{Physical Review E}}
  \textbf{\bibinfo{volume}{87}}, \bibinfo{pages}{062806}
  (\bibinfo{year}{2013}).

\bibitem{menichetti2014weighted}
\bibinfo{author}{Menichetti, G.}, \bibinfo{author}{Remondini, D.},
  \bibinfo{author}{Panzarasa, P.}, \bibinfo{author}{Mondrag{\'o}n, R.~J.} \&
  \bibinfo{author}{Bianconi, G.}
\newblock \bibinfo{title}{Weighted multiplex networks}.
\newblock \emph{\bibinfo{journal}{PloS one}} \textbf{\bibinfo{volume}{9}},
  \bibinfo{pages}{e97857} (\bibinfo{year}{2014}).

\bibitem{cardillo2013emergence}
\bibinfo{author}{Cardillo, A.} \emph{et~al.}
\newblock \bibinfo{title}{Emergence of network features from multiplexity}.
\newblock \emph{\bibinfo{journal}{Scientific Reports}}
  \textbf{\bibinfo{volume}{3}}, \bibinfo{pages}{1344} (\bibinfo{year}{2013}).

\bibitem{braun2015dynamic}
\bibinfo{author}{Braun, U.} \emph{et~al.}
\newblock \bibinfo{title}{Dynamic reconfiguration of frontal brain networks
  during executive cognition in humans}.
\newblock \emph{\bibinfo{journal}{Proceedings of the National Academy of
  Sciences}} \textbf{\bibinfo{volume}{112}}, \bibinfo{pages}{11678--11683}
  (\bibinfo{year}{2015}).

\bibitem{filiposka2017bridging}
\bibinfo{author}{Filiposka, S.}, \bibinfo{author}{Gajduk, A.},
  \bibinfo{author}{Dimitrova, T.} \& \bibinfo{author}{Kocarev, L.}
\newblock \bibinfo{title}{Bridging online and offline social networks:
  Multiplex analysis}.
\newblock \emph{\bibinfo{journal}{Physica A: Statistical Mechanics and its
  Applications}} \textbf{\bibinfo{volume}{471}}, \bibinfo{pages}{825--836}
  (\bibinfo{year}{2017}).

\bibitem{morris2012transport}
\bibinfo{author}{Morris, R.~G.} \& \bibinfo{author}{Barthelemy, M.}
\newblock \bibinfo{title}{Transport on coupled spatial networks}.
\newblock \emph{\bibinfo{journal}{Physical review letters}}
  \textbf{\bibinfo{volume}{109}}, \bibinfo{pages}{128703}
  (\bibinfo{year}{2012}).

\bibitem{sole2016congestion}
\bibinfo{author}{Sol{\'e}-Ribalta, A.}, \bibinfo{author}{G{\'o}mez, S.} \&
  \bibinfo{author}{Arenas, A.}
\newblock \bibinfo{title}{Congestion induced by the structure of multiplex
  networks}.
\newblock \emph{\bibinfo{journal}{Physical review letters}}
  \textbf{\bibinfo{volume}{116}}, \bibinfo{pages}{108701}
  (\bibinfo{year}{2016}).

\bibitem{wang2015coupled}
\bibinfo{author}{Wang, Z.}, \bibinfo{author}{Andrews, M.~A.},
  \bibinfo{author}{Wu, Z.-X.}, \bibinfo{author}{Wang, L.} \&
  \bibinfo{author}{Bauch, C.~T.}
\newblock \bibinfo{title}{Coupled disease--behavior dynamics on complex
  networks: A review}.
\newblock \emph{\bibinfo{journal}{Physics of life reviews}}
  \textbf{\bibinfo{volume}{15}}, \bibinfo{pages}{1--29} (\bibinfo{year}{2015}).

\bibitem{funk2015nine}
\bibinfo{author}{Funk, S.} \emph{et~al.}
\newblock \bibinfo{title}{Nine challenges in incorporating the dynamics of
  behaviour in infectious diseases models}.
\newblock \emph{\bibinfo{journal}{Epidemics}} \textbf{\bibinfo{volume}{10}},
  \bibinfo{pages}{21--25} (\bibinfo{year}{2015}).

\bibitem{granell2013dynamical}
\bibinfo{author}{Granell, C.}, \bibinfo{author}{G{\'o}mez, S.} \&
  \bibinfo{author}{Arenas, A.}
\newblock \bibinfo{title}{Dynamical interplay between awareness and epidemic
  spreading in multiplex networks}.
\newblock \emph{\bibinfo{journal}{Physical review letters}}
  \textbf{\bibinfo{volume}{111}}, \bibinfo{pages}{128701}
  (\bibinfo{year}{2013}).

\bibitem{sanz2014dynamics}
\bibinfo{author}{Sanz, J.}, \bibinfo{author}{Xia, C.-Y.},
  \bibinfo{author}{Meloni, S.} \& \bibinfo{author}{Moreno, Y.}
\newblock \bibinfo{title}{Dynamics of interacting diseases}.
\newblock \emph{\bibinfo{journal}{Physical Review X}}
  \textbf{\bibinfo{volume}{4}}, \bibinfo{pages}{041005} (\bibinfo{year}{2014}).

\bibitem{lima2015disease}
\bibinfo{author}{Lima, A.}, \bibinfo{author}{De~Domenico, M.},
  \bibinfo{author}{Pejovic, V.} \& \bibinfo{author}{Musolesi, M.}
\newblock \bibinfo{title}{Disease containment strategies based on mobility and
  information dissemination}.
\newblock \emph{\bibinfo{journal}{Scientific reports}}
  \textbf{\bibinfo{volume}{5}}, \bibinfo{pages}{10650} (\bibinfo{year}{2015}).

\bibitem{holme2002growing}
\bibinfo{author}{Holme, P.} \& \bibinfo{author}{Kim, B.~J.}
\newblock \bibinfo{title}{Growing scale-free networks with tunable clustering}.
\newblock \emph{\bibinfo{journal}{Physical review E}}
  \textbf{\bibinfo{volume}{65}}, \bibinfo{pages}{026107}
  (\bibinfo{year}{2002}).

\bibitem{feenstra2000nber}
\bibinfo{author}{Feenstra, R.} \& \bibinfo{author}{Lipsey, R.}
\newblock \bibinfo{title}{Nber-united nations trade data 1962-2000}
  (\bibinfo{year}{2000}).

\bibitem{banerjee2013diffusion}
\bibinfo{author}{Banerjee, A.}, \bibinfo{author}{Chandrasekhar, A.~G.},
  \bibinfo{author}{Duflo, E.} \& \bibinfo{author}{Jackson, M.~O.}
\newblock \bibinfo{title}{The diffusion of microfinance}.
\newblock \emph{\bibinfo{journal}{Science}} \textbf{\bibinfo{volume}{341}},
  \bibinfo{pages}{1236498} (\bibinfo{year}{2013}).

\bibitem{hidalgo2007product}
\bibinfo{author}{Hidalgo, C.~A.}, \bibinfo{author}{Klinger, B.},
  \bibinfo{author}{Barab{\'a}si, A.-L.} \& \bibinfo{author}{Hausmann, R.}
\newblock \bibinfo{title}{The product space conditions the development of
  nations}.
\newblock \emph{\bibinfo{journal}{Science}} \textbf{\bibinfo{volume}{317}},
  \bibinfo{pages}{482--487} (\bibinfo{year}{2007}).

\bibitem{hidalgo2009building}
\bibinfo{author}{Hidalgo, C.~A.} \& \bibinfo{author}{Hausmann, R.}
\newblock \bibinfo{title}{The building blocks of economic complexity}.
\newblock \emph{\bibinfo{journal}{proceedings of the national academy of
  sciences}} \textbf{\bibinfo{volume}{106}}, \bibinfo{pages}{10570--10575}
  (\bibinfo{year}{2009}).

\bibitem{tacchella2012new}
\bibinfo{author}{Tacchella, A.}, \bibinfo{author}{Cristelli, M.},
  \bibinfo{author}{Caldarelli, G.}, \bibinfo{author}{Gabrielli, A.} \&
  \bibinfo{author}{Pietronero, L.}
\newblock \bibinfo{title}{A new metrics for countries' fitness and products'
  complexity}.
\newblock \emph{\bibinfo{journal}{Scientific reports}}
  \textbf{\bibinfo{volume}{2}}, \bibinfo{pages}{723} (\bibinfo{year}{2012}).

\bibitem{cristelli2013measuring}
\bibinfo{author}{Cristelli, M.}, \bibinfo{author}{Gabrielli, A.},
  \bibinfo{author}{Tacchella, A.}, \bibinfo{author}{Caldarelli, G.} \&
  \bibinfo{author}{Pietronero, L.}
\newblock \bibinfo{title}{Measuring the intangibles: A metrics for the economic
  complexity of countries and products}.
\newblock \emph{\bibinfo{journal}{PloS one}} \textbf{\bibinfo{volume}{8}},
  \bibinfo{pages}{e70726} (\bibinfo{year}{2013}).

\bibitem{stojkoski2016impact}
\bibinfo{author}{Stojkoski, V.}, \bibinfo{author}{Utkovski, Z.} \&
  \bibinfo{author}{Kocarev, L.}
\newblock \bibinfo{title}{The impact of services on economic complexity:
  Service sophistication as route for economic growth}.
\newblock \emph{\bibinfo{journal}{PloS one}} \textbf{\bibinfo{volume}{11}},
  \bibinfo{pages}{e0161633} (\bibinfo{year}{2016}).

\bibitem{Granovetter1973}
\bibinfo{author}{Granovetter, M.~S.}
\newblock \bibinfo{title}{The strength of weak ties}.
\newblock \emph{\bibinfo{journal}{American journal of sociology}}
  \bibinfo{pages}{1360--1380} (\bibinfo{year}{1973}).

\bibitem{holes1992social}
\bibinfo{author}{Holes, S.}
\newblock \bibinfo{title}{The social structure of competition}
  (\bibinfo{year}{1992}).

\end{thebibliography}



\section*{Author contributions statement}
All authors conceived the experiments.  T.D. and K.P. conducted the experiments. All authors analyzed the results, wrote and reviewed the manuscript. 

\section*{Additional information}
\textbf{Competing financial interests} The authors declare no conflicts of interest. 

\end{document}